# PROOF as a Service on the Cloud: a Virtual Analysis Facility based on the CernVM ecosystem

**D Berzano, J Blomer, P Buncic, I Charalampidis, G Ganis, G Lestaris, R Meusel**
CERN PH-SFT
CERN, CH-1211 Geneva 23, Switzerland

E-mail: dario.berzano@cern.ch

**Abstract.** PROOF, the Parallel ROOT Facility, is a ROOT-based framework which enables interactive parallelism for event-based tasks on a cluster of computing nodes. Although PROOF can be used simply from within a ROOT session with no additional requirements, deploying and configuring a PROOF cluster used to be not as straightforward. Recently great efforts have been spent to make the provisioning of generic PROOF analysis facilities with zero configuration, with the added advantages of positively affecting both stability and scalability, making the deployment operations feasible even for the end user. Since a growing amount of large-scale computing resources are nowadays made available by Cloud providers in a virtualized form, we have developed the Virtual PROOF-based Analysis Facility: a cluster appliance combining the solid CernVM ecosystem and PoD (PROOF on Demand), ready to be deployed on the Cloud and leveraging some peculiar Cloud features such as elasticity. We will show how this approach is effective both for sysadmins, who will have little or no configuration to do to run it on their Clouds, and for the end users, who are ultimately in full control of their PROOF cluster and can even easily restart it by themselves in the unfortunate event of a major failure. We will also show how elasticity leads to a more optimal and uniform usage of Cloud resources.

## 1. Introduction

The adoption of virtualization technologies in the computing of High Energy Physics has rapidly increased over the last years: presently, many small to large Grid "tiers" are running Grid jobs transparently inside virtual machines.

Virtualization offers a distinct separation between "computing resources" and "services" in a way that goes beyond merely supporting the Grid use case: by exposing an interface where users can submit virtual machines with a limited lifetime instead of Grid jobs, the creation of personal virtual clusters becomes practicable, effectively implementing a model to which we can refer as "Tier-3s on demand".

Such virtual clusters might come in the form of a preconfigured service—an "appliance"—ready to be executed on the cloud: the work we present is a Virtual Analysis Facility based on PROOF [1], the parallel computing framework integrated with ROOT [2]; all the necessary tools are in practice configured by contextualizing the virtual machines of the cluster. The resulting appliance is a ready to go "PROOF as a Service", configurable by end users in a few clicks through a web interface: users will not even need to login to the created virtual machines as no further configuration step is needed.

In section 2 we explain how we have contributed to PROOF by making it compliant with a cloud-based analysis model. Section 3 describes the components of the Virtual Analysis Facility, and in particular its CernVM [3] foundations; a component called Elastiq, capable of expanding and reducing the size of our virtual cluster automatically, is covered in section 3.2. Finally, in section 4 we present some performance considerations and measures showing how PROOF leads to analysis results faster than an equivalent set of Grid jobs (section 4.1); benchmarks showing how rapidly the Virtual Analysis Facility scales when needed are also reported (section 4.2).

## 2. PROOF as a cloud-aware analysis model

When doing high-performance computing on the cloud no tight assumptions can be made regarding the quality and amount of resources our application will get. Cloud resources are heterogeneous by definition: two virtual machines configured to have the same resources (RAM, CPU, disk) might perform very differently in practice. Moreover, virtual machines are volatile: they can appear and disappear at any time, and we might never get the number of virtual machines we have requested.

We call "cloud-aware" a virtual appliance capable of dealing smoothly with changes in the underlying resources. For what concerns HEP appliances exploiting event-based parallelism, cloud-awareness is attainable by performing no prior pinning of data to the computing resources: from this perspective, the Grid batch model is not cloud-aware, as the input dataset is pre-split to the submitted jobs, and we can obtain the results only when all of them have executed.

The PROOF computing model is interactive: a "master" process maintains constant communication with its distributed "workers". Interactive communications make PROOF cloud-aware by sending over pointers to data to analyse dynamically, and by allowing new workers to join and offload a currently running process, as has been finally implemented in recent developments. PROOF's cloud-aware components as well as its new features are illustrated in the next paragraphs.

### *2.1. PROOF on Demand*

PROOF on Demand (PoD) [4] is a toolchain capable of running PROOF without the need of static multiuser installations. PoD works either opportunistically, via SSH, or by submitting PROOF workers to a workload management system. When using PoD, users have full control over their personal PROOF sandbox, meaning that any potential problem will not affect other users and will not require any system administrator intervention, since users can restart their PROOF sandbox themselves.

PoD also runs on the Grid: the ATLAS collaboration is currently testing an experimental PoD plugin for the PanDA middleware [5].

### *2.2. Adaptive packetizer*

PROOF features a pull scheduler (the "packetizer") to assign data ("packets") to workers. The master manages the dataset and the list of workers: whenever a worker has finished processing a packet it requests a new one, until all the packets have been processed. Since PROOF is closely integrated with ROOT, the packetizer understands ROOT's file format and knows how to split such files in smaller packets. In cases where input files are distributed on the computing nodes themselves, the packetizer always favours local processing. By assigning the workload non-uniformly and granularly, PROOF ensures that the completion time of all workers is as uniform as possible [6].

### *2.3. Dynamic workers addition*

PROOF workers can be started dynamically and added to a currently running process: all the initialization process (loading libraries, packages and analysis macros) is replayed on the newly added workers, which are subsequently taken into consideration by the packetizer.

This new feature we implemented starting from ROOT v5.34.10 makes PROOF deal optimally with cases where resources become gradually available after their request: this occurs for instance either with PROOF on Demand over batch systems—where PROOF workers are queued altogether and started when possible, or with our Virtual Analysis Facility—where queuing PROOF workers triggers an automatic background request for instantiating new virtual machines.

Implementing such feature has a direct impact on the user's workflow, who does not need to wait for a certain number of workers to be available before starting her analysis. This is useful because the num-ber of obtained workers, with respect to the requested ones, is uncertain: with the dynamic addition of workers, an analysis can start when the first worker is available, others will join seamlessly as soon as they can.

From the technical point of view, dynamic addition of workers involves several components. When a new worker arrives, PROOF on Demand updates a list of available workers shared with the PROOF master, which in turn periodically polls such list: as soon as new workers are found, they are initialized as described and communicated to the packetizer. The reason why polling occurs periodically is to aggregate the initialization of new workers as much as possible, which occurs in parallel.

Such feature leads to a more optimal utilization of resources: since users do not have to worry anymore of having "enough" workers before starting the analysis, the initial user wait time is completely eliminated. Moreover, workers that are started when PROOF is processing are not left idle anymore, because they are now used in the current process.

In section 4.1 we will demonstrate analytically how such dynamic workflow applied to PROOF resources obtained on the Grid outperforms the same batch analysis split into independent batch jobs on the same set of resources.

## 3. A cloud-aware Virtual Analysis Facility

The Virtual Analysis Facility is practically a set of $\mu$CernVM [7] virtual machines—one head node plus a variable number of workers, configured to work together as a HTCondor batch cluster. PROOF, the central component of the computing model we propose, is controlled through PROOF on Demand: the PROOF master, one for each user, is deployed on the head node, while PROOF workers are submitted as HTCondor jobs.

In the following paragraphs we describe the components which constitute the Virtual Analysis Facility: in particular we will cover the configuration procedure through CernVM Online (section 3.1.1) and the capability to scale automatically by means of the new Elastiq daemon (section 3.2).

### *3.1. The CernVM ecosystem*

The base virtual machine used in the Virtual Analysis Facility is $\mu$CernVM, a SLC6-based operating system entirely downloaded and cached on demand via CernVM-FS [8], which is used to serve the entire root filesystem. Being CernVM-FS a read-only filesystem, a writable local partition is overlaid, making it usable as if it was entirely local in a completely transparent way.

As all operating system files are downloaded when needed, the base image is extremely lightweight and fast to deploy (less than 15 MB), yet it features a fully fledged installation, as opposed to a stripped down "just-enough" operating system.

CernVM-FS is also used to provide access to specific experiment maintained software repositories.

#### *3.1.1. CernVM Online*

The CernVM philosophy is to maintain a single generic base image and to configure it at boot time via an appropriate contextualization, as opposed to snapshotting a configured image.

The CernVM ecosystem includes CernVM Online [9], a simple web interface to create and store virtual machine contexts securely. An even simpler interface is presented to the user when creating a context for the Virtual Analysis Facility, as we can see in figure 1: the final result will be a configuration text file that we can download and pass as "user-data" to our cloud controller when instantiating the $\mu$CernVM base image.

Once started, the Virtual Analysis Facility is fully configured and can be used right away: it is not even needed to log in to perform further configuration. Great care has been put in making the configuration process as simple as possible, to allow end users with no knowledge of system administration to start their own PROOF cluster almost effortlessly.

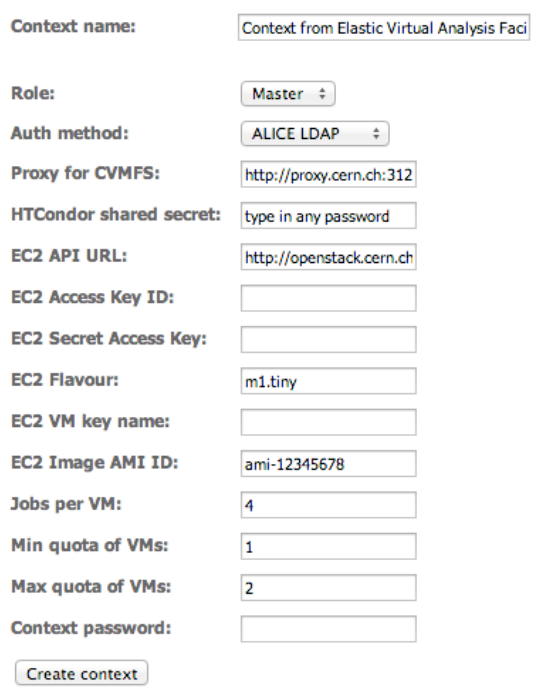

**Figure 1.** The extremely simple configuration interface of the Virtual Analysis Facility from the CernVM Online web site.

### *3.2. HTCondor*
HTCondor [10] is a popular open source workload management system. HTCondor allows for cluster worker nodes to autoregister and deregister, whereas alternatives such as TORQUE [11], used in an early prototype of the Virtual Analysis Facility [12], need a manual update of the nodes list.

### *3.3. Elastiq queue monitor*
The Virtual Analysis Facility is capable of scaling up and down automatically based on the current resources usage. To allow for such functionality we developed a Python daemon called Elastiq [13].
Elastiq runs in background to monitor both the HTCondor jobs queue and the number of jobs running on each HTCondor node. PROOF workers requested via PROOF on Demand end up in the jobs queue: when there are too many jobs waiting for too long, Elastiq requests a certain number of new virtual machines to the cloud hypervisor: jobs threshold, waiting time and number of virtual machines per waiting job can all be configured. Since all new HTCondor nodes will autoregister when ready as of section 3.2, Elastiq has no additional step to perform. On the other hand, when some virtual machines appear to be idle for too long, Elastiq requests their shutdown; idle time can also be configured, as well as a quota of minimum and maximum number of running virtual machines.
Communication with the cloud hypervisor occurs through the EC2 API using three parameters (API URL, access key ID and secret key) configured during the contextualization as visible in figure 1.
Elastiq performs the requests by intentionally ignoring EC2 API errors (which might occur for instance when the user's cloud quota has been exceeded), as the requests will be periodically reissued if still necessary.
Future developments of Elastiq include support for the CernVM Cloud meta cloud controller API [14].

### *3.4. Security and authentication*
The Virtual Analysis Facility is accessed using the PoD client, which relies on SSH for authentication, encryption of communications and tunnelling of the PROOF connection. Passwordless SSH access to the head node is required. In case user has a X.509 certificate and key (all Grid users do), the client can perform a two-step authentication using HTTPS and SSH using a standard web browser and SSH client thanks to sshcertauth [15].

## 4. Performance measures and considerations
Since the Virtual Analysis Facility is constituted of a variety of different components, we have tested PROOF alone and the rest of the Virtual Analysis Facility setup separately. In the first test, based on actual startup latencies, we have compared the PROOF and the batch computing model on the same set of Grid resources. The second test shows the responsiveness of the virtual cluster setup by measuring the time spent between a user's workers request and their actual startup, where the request involves the creation of virtual machines from scratch.

### *4.1. Comparison of PROOF and Grid computing models*
The combination of the adaptive workload assignment and the new dynamic workers addition feature is what makes PROOF a cloud-aware computing model based on pull scheduling. In addition, by means of PROOF on Demand, PROOF workers can be submitted as if they were Grid jobs. Within the context of the ATLAS experiment, we have measured on several Grid sites the startup latencies of 100 Grid jobs (or PROOF workers): submissions have been done through the experimental PoD PanDA plugin [5]. The plot represented in figure 2 shows the ramp-up of the running Grid jobs within a time window of one hour: such results are the starting point for the following analytical comparison between the Grid and the PROOF computing models.

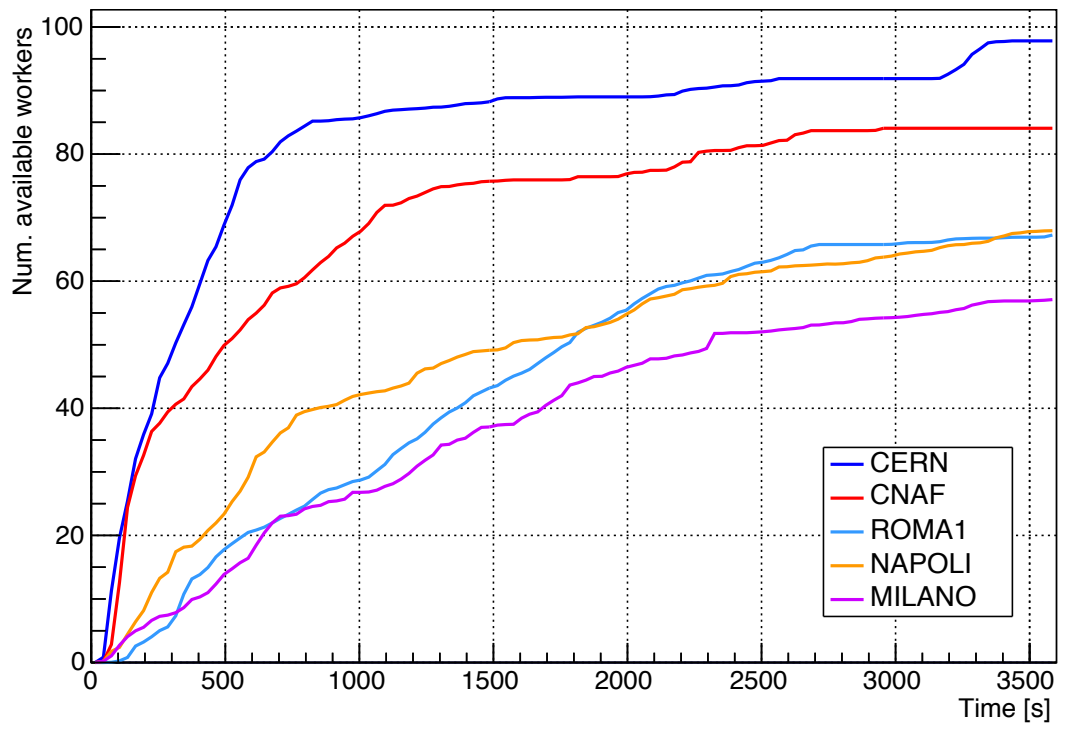

**Figure 2.** Ramp-up of running Grid jobs in a time window of one hour after submitting 100 jobs on different sites.

Data shown in figure 2 can be fitted using a simple function representing the number of running jobs as a function of time:

$$n(t) = \frac{p_0 t}{1 + p_1 t} \tag{1}$$

with $t$ being the time elapsed after the first job is started. Parameters $p_0$ and $p_1$ are representative of each Grid site, and in particular the ratio $p_0/p_1$ is the maximum number of jobs per user running on a single site. The time needed to obtain $n$ running jobs can be obtained by inverting (1):

$$t(n) = \frac{n}{p_0 - p_1 n} \tag{2}$$

If jobs are PROOF workers, thanks to the dynamic workers feature all of the jobs running time is spent in processing, and workers are never idle. If we assume that the initialization time for a worker is negligible, a PROOF analysis completing after a time $t'$ uses a serialized computing time given by:

$$T(t') = \int_0^{t'} n(t)\,dt = \frac{p_0}{p_1^2}\left(p_1 t' - \log(1 + p_1 t')\right) \tag{3}$$

We now consider the pure push scheduling case, where Grid jobs are independent. Since the number of jobs is fixed and data is pinned to each job, we must wait for the last started job to complete before the whole analysis has finished. If our reference analysis requires a total serialized time $T$ and it is divided into $n'$ jobs, it will complete after a time $t'$ (the "time to results") given by:

$$t' = t(n') + \frac{T}{n'} = \frac{n'}{p_0 - p_1 n'} + \frac{T}{n'} \tag{4}$$

where the first term is the time elapsed before the last ($n'$-th) worker is available, derived from (2), and the second term is its processing time, assumed to be even for all jobs.
The dependency from $n'$ can be eliminated if we choose it as the "optimal" value, *i.e.* the one that minimizes $t'$: from the zeroes of the derivative of (4) we obtain an expression for $n'$ which depends on the serialized processing time $T$:

$$n'(T) = \frac{p_0\sqrt{T}}{\sqrt{p_0} + p_1\sqrt{T}} \quad (5)$$

We now substitute (5) into (4):

$$t'_{push}(T) = \frac{2\sqrt{p_0 T} + p_1 T}{p_0} \quad (6)$$

The function (6) is plotted along with the numerically inverted (3) in figure 3, and their ratio is represented in figure 4; for both plots the choice of $p_0$ and $p_1$ comes from the CERN site.

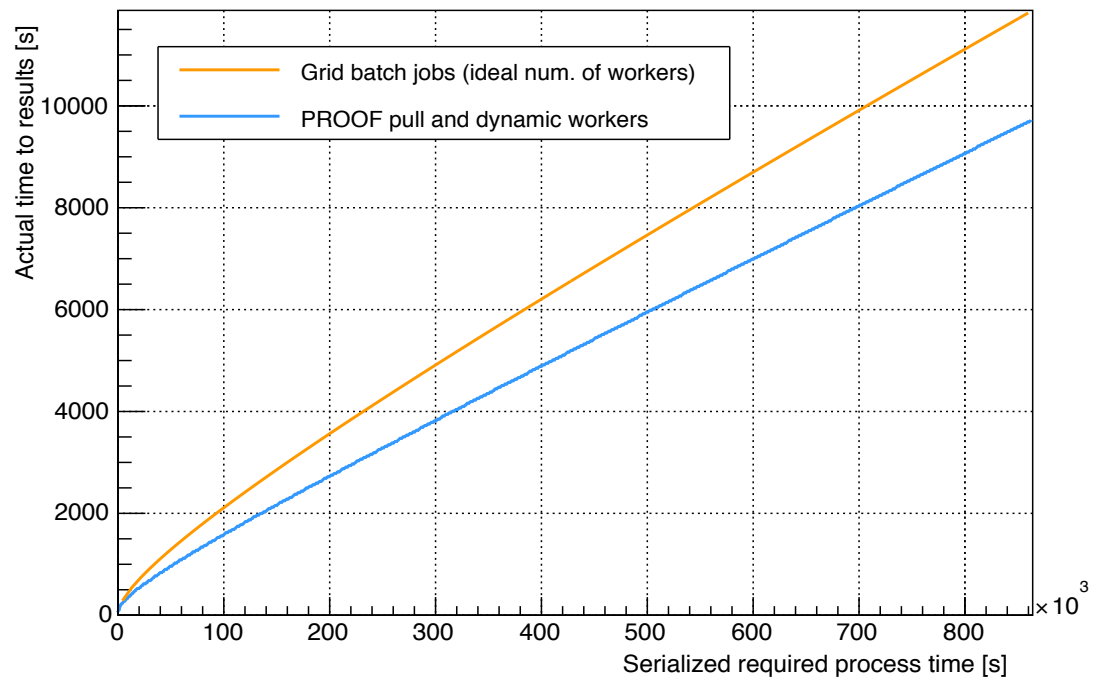

**Figure 3.** Push and pull scheduling time to results vs. the serialized computing time.

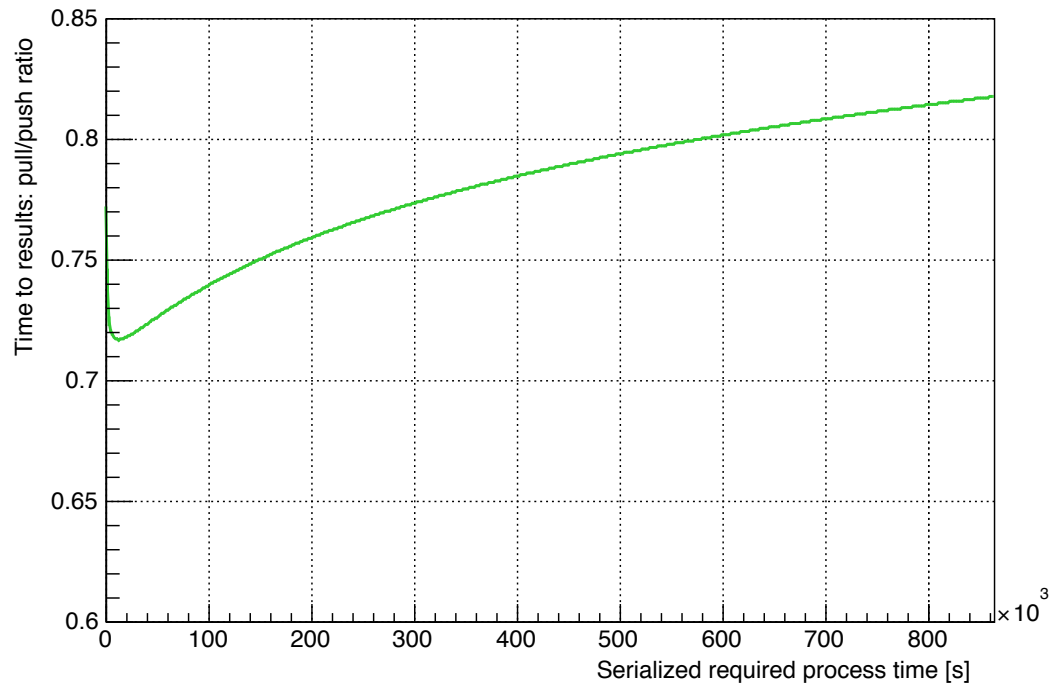

**Figure 4.** Pull/push time to results ratio for serialized computing times up to 10 days.

Although pull and push scheduling are asymptotically equivalent, for realistic serialized job processing times, the PROOF pull model always yields a considerable speedup: by using again parameter values from the CERN site, a process requiring a serialized computing time of 10 days takes 2 hr 42 min to complete while the same process as Grid jobs would take 3 hr 18 min, with PROOF being 18% faster in producing the results. Shorter analyses, which are the common use case with PROOF, benefit from an even greater speedup: an analysis requiring two days of total computing time (40 min with PROOF, 53 min with Grid jobs) is 25% faster on PROOF. Measurements of the actual required analysis time of PROOF on the Grid to corroborate this analytical comparison are under way.

### *4.2. Startup latency of μCernVM and PROOF*

Given the Virtual Analysis Facility feature of scaling up by requesting new virtual machines when needed, we are interested in measuring the delay before the requested PROOF workers join the PROOF cluster by starting from scratch, *i.e.* with the sole head node virtual machine running.

The sequence of background operations after requesting the workers that contribute to the latency is: virtual machine deployment, μCernVM boot, HTCondor node registration and PROOF reaction time.

The virtual machine deployment depends on several factors, notably the base image size and the resources availability. As we have seen in section 3.1 the μCernVM base image size is so small that it may be safely considered negligible; we have also made sure that enough resources were available to start all the requested virtual machines immediately.

Since μCernVM downloads the necessary files on demand via HTTP requests performed by CernVM-FS, a local caching proxy server is normally configured to avoid duplicated downloads. To measure a boot time under similar conditions, we have pre-run before our tests a dummy μCernVM virtual machine merely to allow the local HTTP proxy to cache all data needed at boot.

Tests were run on two cloud infrastructures: the CERN Agile infrastructure [16], based on OpenStack, and the INFN Torino private cloud [17], based on OpenNebula. In both cases the number of submitted PROOF workers was enough to trigger the deployment of 10 new virtual machines.

Results were collected via the “pod-info -l” command, which shows the latency for each PROOF worker. Average values are compatible: 6 min 15 s ± 39 s for the CERN cloud and 5 min 51 s ± 21 s for the Torino cloud—yielding an average latency of around 6 minutes, which is a perfectly acceptable waiting time, especially if we consider that the started virtual machines will be reused many times.

## 5. Conclusions

The Virtual Analysis Facility has been inspired by the principle of reusing as much as possible existing solid and independent components, and by favouring an appropriate setup of such tools over the development of new ones. We have seen how such selected components are cloud-aware: both PROOF and HTCondor support dynamic addition of workers, while $\mu$CernVM and CernVM-FS retrieve all software and system components only when needed exposing a standard FUSE mount point.
Development has been necessary to implement the dynamic workers addition in PROOF, which finally offers functionalities similar to other non-HEP specific tools like Hadoop. As tests have shown [18], PROOF still performs 20% better than Hadoop on the same dataset, essentially because it benefits from the knowledge of the input data format, allowing for a fine grained data assignment, as seen in section 2.2. The implementation of such feature is not restricted to the Virtual Analysis Facility, being the ATLAS experiment with PoD over PanDA the first potential customer.
The Elastiq daemon has also been created to provide “elasticity” to the Virtual Analysis Facility, but it can be also installed separately to scale up and down any HTCondor virtual cluster via the EC2 API.
By hiding the complexity of the components behind a simple configuration interface as seen in section 3.1.1 the Virtual Analysis Facility is an effective implementation of a Tier-3 on demand.